\documentclass[aps,prd,twocolumn,nofootinbib,longbibliography,10pt]{revtex4-2}
\usepackage{etoolbox}
\usepackage{dcolumn,tensor,nicefrac}
\usepackage{amsmath,amssymb,amsfonts,mathtools}
\usepackage{mathrsfs,bbold}
\usepackage{graphicx}
\usepackage{subcaption}
\usepackage[colorlinks=true,urlcolor=blue,citecolor=red,linkcolor=blue]{hyperref}
\usepackage{accents}
\newlength{\dhatheight}
\usepackage{natbib}
\usepackage{wrapfig}
\usepackage{booktabs}
\usepackage{float}
\usepackage{flushend,BOONDOX-cal,BOONDOX-frak}

\makeatletter
\newsavebox{\@brx}
\newcommand{\llangle}[1][]{\savebox{\@brx}{\(\m@th{#1\langle}\)}%
  \mathopen{\copy\@brx\kern-0.5\wd\@brx\usebox{\@brx}}}
\newcommand{\rrangle}[1][]{\savebox{\@brx}{\(\m@th{#1\rangle}\)}%
  \mathclose{\copy\@brx\kern-0.5\wd\@brx\usebox{\@brx}}}
\makeatother

\begin{document}
\title{\textbf{Spin-curvature effects in slowly rotating black hole spacetime}}
\author{Arpita Jana}
\email{janaarpita2001@gmail.com}
\affiliation{Department of Astrophysics and High Energy Physics, S. N. Bose National Centre for Basic Sciences, JD Block, Sector-III, Salt Lake City, Kolkata-700 106, India}
\author{Subhajit Mazumdar}
\email{subhajitmazumdar@gmail.com}
\affiliation{Department of Astrophysics and High Energy Physics, S. N. Bose National Centre for Basic Sciences, JD Block, Sector-III, Salt Lake City, Kolkata-700 106, India}
\author{Sunandan Gangopadhyay}
\email{sunandan.gangopadhyay@gmail.com}
\affiliation{Department of Astrophysics and High Energy Physics, S. N. Bose National Centre for Basic Sciences, JD Block, Sector-III, Salt Lake City, Kolkata-700 106, India}

\begin{abstract}
\noindent We investigate the spin-curvature interaction for a massive spin-1/2 quantum particle in a slowly rotating black hole spacetime. In contrast to the flat Minkowski space, the curved spacetime introduces non-trivial spin connections into the Dirac equation due to the direct curvature-spin interaction. The force arising from this non-trivial spin-curvature coupling makes the particle deviate from its geodesics. 
We first calculate the tetrads in the co-moving frame of the particle in the slowly rotating black hole background.
We then compute the components of the relativistic quantum force, arising from the spin-curvature interaction, using these co-moving tetrads up to linear order in the rotation parameter ($a$) of the black hole. The procedure of computing the force is based on the WKB approximation and the Gordon decomposition method for the Dirac probability four current. Our results show that the rotation parameter modifies the magnitude of the force  by introducing a term that decays more rapidly with radial distance than the force magnitude in a static spherically symmetric Schwarzschild black hole spacetime. 
 
\end{abstract}
\maketitle
\section{Introduction}
\noindent 
 Quantum field theory in curved spacetime brings together two pillars of theoretical physics in one platform, namely, general theory of relativity and quantum mechanics, and this remains a vibrant field of active theoretical research  \cite{birrelldavis, fronsdal1,fronsdal2,mukhanov,wald}. Unlike the scalar and vector quantum fields, which can be treated by the application of minimal substitution rule following from the general covariance principle, extending this framework to massive spin-1/2 quantum fields is quite non-trivial, and reveals richer and interesting physical phenomena. In the middle of the twentieth century, Mathisson \cite{mathisson} and Papapetrou \cite{papapetrou} described the behaviour of spinning particles in curved spacetime, and in their analysis, they performed a multipole expansion around the worldline and showed that the motion turns out to be non-geodesic due to the interaction between the spin of the particle and the curvature of the spacetime geometry. Later, in \cite{audretsch}, the semi-classical WKB expansion method was applied to massive Dirac fields in arbitrary curved spacetime geometry, and the spin-curvature interaction affects were shown to affect the trajectory of the quantum spin-1/2 particle. The case of non-vanishing torsion has been discussed in \cite{audretsch2}. \\
 \noindent The intrinsic spin of the Dirac field quanta and the curvature of the spacetime geometry, gives rise to non-trivial physics. In particular, a force due to the spin-curvature interaction term arises in the analysis. Understanding this relativistic quantum force turns out to be crucial for several foundational and practical reasons. At $\mathcal{O}(\hbar)$, the interaction between the intrinsic spin of the quantum Dirac particle and curvature of the spacetime geometry, forces the particle
to deviate from the classical geodesic trajectory, thereby providing a more accurate picture of how matter behaves when gravitational effects are taken into account in quantum mechanics. Since the particle does not follow a geodesic trajectory, a natural question that arises is whether the equivalence principle is violated or not. This is one of the fundamental reasons which makes the study of the motion of spin-1/2 quantum particles in curved spacetime geometry, in the presence of these relativistic quantum forces generating from the interaction between spin of the quantum particle and spacetime curvature, extremely important. The weak equivalence principle states that all structureless test particles will fall with exactly the same acceleration in a given gravitational field, independent of their masses, which in turn implies that they follow the same geodesic trajectory. So one may conclude from the non-geodesic motion of the particles that there is indeed a violation of the equivalence principle, which forms the backbone of general theory of relativity. 
However, there is inherently no violation of Einstein's equivalence principle 
due to the deviation of the paths of the quantum spin-1/2 particles from their geodesic trajectories
since quantum particles are described by wave packets which spread out over a finite region of space. In a curved spacetime geometry, this spreading out of the wave packet of the quantum particle play an important role as the interaction due to the spin-curvature coupling samples the spatial extension of the wave packet in the curved spacetime geometry at multiple points simultaneously. Consequently, quantum particles with intrinsic spin, due to the spatial extension of their wave packet, feel a tidal force arising from the interaction of spin and curvature  \cite{semerak,felice,zakharov,geroch1,clarke,isham,milnor,geroch2,hawking,avis,rudiger}, and as an effect of this force, the motion of the spinning quantum particles deviates from the geodesics \cite{corinaldesi,papapetrou,plyatsko1,plyatsko2,obukhov}. \\ 
\noindent The study of spinning quantum particles in a rotating spacetime geometry is a natural extension of the works mentioned in the above references. Since astrophysical objects do have a rotation, this motivates us to carry out the investigation in a slowly rotating spacetime geometry, which brings in additional effects into the picture due to the rotating geometry. In particular, the calculation of the relativistic force arising from the spin-curvature interaction in a rotating spacetime geometry, is something which is worth carrying out, and has been missing in the literature. We now mention some of the works that have been carried out in a rotating spacetime geometry. In \cite{lanzagortasalgado}, the authors studied the effect of gravitational frame dragging on the spin-1/2 quantum particle propagating in curved spacetime. They considered the motion of the massive spin-1/2 quantum particle in the Kerr geometry along the geodesic, without taking into account the geodesic deviation arising from the spin-curvature coupling force. In the present study, we have considered the $\mathcal{O}(\hbar)$ correction to the geodesic four velocity of the spin-1/2 quantum particle propagating in a slowly rotating black hole spacetime and calculated the spin-curvature coupling force for equatorial circular orbits.\\
\noindent Another important motivation for understanding this relativistic quantum force, arising from spin-curvature interaction, come from the practical need of protecting quantum information systems. Gravity acts as a source of environmental noise and therefore it becomes absolutely essential
to understand the dynamics of quantum information carrier in a gravitational field. For this we need a framework which brings together relativistic quantum mechanics and curved spacetime in one platform. As has been mentioned earlier, the propagation of quantum information in curved spacetime is commonly studied by treating the information carrier following a classical geodesic trajectory together with the evolution of its internal spin degrees of freedom. Although this description captures the leading semiclassical 
behaviour, it ignores the quantum correction arising from the coupling of the intrinsic quantum spin to the background spacetime curvature. At $\mathcal{O}(\hbar)$, the spin-curvature coupling force induces a deviation from the geodesic path and provides a direct measure of the effect of coupling of intrinsic quantum spin and spacetime curvature on the dynamics of a relativistic quantum spin-$1/2$ particle. A slowly rotating black hole spacetime provides a natural framework for investigating this effect because in this gravitational background, the rotation induces frame dragging in addition to spacetime curvature. Therefore, the calculation of the spin-curvature force on a Dirac particle in a slowly rotating black hole background provides the understanding how the rotating gravitational background modifies the trajectory of a relativistic spin-$1/2$ quantum particle, and determines the limits in which the classical geodesic approximation taken in earlier works remains valid or quantum corrections are significant enough to be considered. This investigation therefore provides us the framework for understanding the dynamics of quantum information carriers moving in background gravitational fields from rotating spacetime geometry and opens the road for analyzing  relativistic quantum information effects.\\
\noindent Before moving further, we would like to mention that for the classical spinning particles, identifying their four-momentum with the convection current, gives the equations for their trajectories, and time evolution of their spins can be obtained using Papapetrou formulations \cite{papapetrou}. In the quantum mechanical scenario, the coupling of the intrinsic quantum spin of the particle with the gravitational field arising from spacetime curvature of the background geometry affects the trajectories of the spinning quantum particles. 
There is however a subtle point that one needs to take care of. Due to the negative energy solutions of the Dirac equation, it is difficult to relate the dynamical variables of the Dirac theory to the corresponding classical analysis. So, one needs to restrict the influence of negative energy solutions to establish the correspondence with classical mechanics. The most acceptable way of doing this is the WKB approximation in an observer independent way. The wave function of the particle then can locally be expressed as \cite{audretsch, lanzagorta}
\begin{equation}\label{diracansatz}
\Psi(x)=e^{i \mathcal{S}(x)/\hbar}\sum_{n=0}^{\infty}(-i\hbar)^{n}\psi_{n}(x)
\end{equation}
with a rapidly varying phase $\mathcal{S}(x)$. As discussed in \cite{kristian, ehlers}, in general relativity, the transition from wave optics to geometrical optics can be executed using such ansatz. In 1932, Pauli discussed the classical limit of the special relativistic Dirac equation in an electromagnetic field \cite{pauli}. After that, an extended analysis has been done in 1963 \cite{rubinow}, which showed that the spinor amplitudes $\psi_n(x)$ satisfy a system of first-order differential equations along the relativistic trajectories.
In this article, we have studied the effect of spacetime curvature-spin interaction for a massive spin-1/2 particle in a slowly rotating black hole spacetime. The curved spacetime geometry introduces a non-trivial spin connection in the Dirac equation, as a result of the interaction between the intrinsic spin of the quantum particle and the curvature of the black hole spacetime. The relativistic quantum force arising from this non-trivial coupling makes the spin-1/2 particles to deviate from the geodesics. We have computed the relativistic quantum force due to the spin-curvature coupling in the co-moving frame and compared the results with the Schwarzschild black hole case, discussed in \cite{lanzagorta}. We found that the rotating spacetime geometry modifies the magnitude of the force
and generates a term that decays with radial distance more rapidly than the magnitude of the force in the non-rotating case.\\
\noindent The article is arranged as follows. In section \ref{tetrad}, we have revisited the tetrad formalism and written down the tetrad basis for the slowly rotating black hole spacetime, followed by the tetrads computed in the co-moving frame for the same spacetime geometry in section \ref{cotetrad}. In section \ref{forcegeneral}, we have presented a review on the description of Dirac spinors in curved spacetime, using the WKB approximation. We have then calculated the relativistic force acting on the quantum particle due to spin-curvature coupling up to linear order in $\hbar$. The first term of the force expression shows the direct interaction of the spacetime curvature and the intrinsic spin of the quantum particle, while the second term is linearly dependent on the mass of the spin-1/2 quantum particle. In section \ref{forcerotating}, we have calculated the magnitude of the force in the co-moving frame for the slowly rotating black hole spacetime. We have done this calculation only for the spin-curvature interaction term, that is, the first term appearing in the expression of the relativistic quantum force, up to linear order in the rotation parameter ($a$). Then we have considered a special case by taking a circular orbit where the radial component of the four velocity of the quantum particle vanishes, and calculated the force magnitude. In section \ref{conclusion}, we have discussed our results and make important observations. In the Appendix, we have presented the components of the connection one-forms for slowly rotating black hole spacetime in the co-moving frame.

\section{The tetrad fields for slowly rotating spacetime}\label{tetrad}
\noindent The most convenient way to study the dynamics of massive spin-1/2 elementary particles in gravitational fields is using the local inertial frames defined at each point of the spacetime. The tetrad field is a set of four linearly independent vector fields \cite{wald, felice,zakharov,lanzagorta} in this local inertial frame \footnote{The Latin indices $a,b,c,...$, are for the local inertial frame, and the Greek indices $\mu,\nu,\lambda,...$, are for the general coordinate system.}. A four-vector field $\Theta^{\mu}$ in the general coordinate system can be expressed in a local inertial frame as 
\begin{equation}
    \Theta^{\mu}(x)=e_{a}^{~\mu}(x) \Theta^{a}(x)~.
\end{equation}
The spacetime metric $g^{\mu \nu}(x)$ is related to the Minkowski metric $\eta^{ab}$ through the tetrad fields as\footnote{The metric tensor in flat spacetime is taken to be mostly positive, $\eta_{ab}=diag(-1,1,1,1)$. } 
\begin{align}\label{tetradprop1}
    & g^{\mu \nu}(x)=e_{a}^{~\mu}(x) e_{b}^{~\nu}(x) \eta^{ab}, \nonumber \\ &
    \eta^{ab}=e^{a}_{~\mu}(x) e^{b}_{~\nu}(x) g^{\mu \nu}(x)
\end{align}
with the orthonormality conditions of the tetrads
\begin{align}\label{tetradprop2}
    &e^{a}_{~\mu}(x) e_{a}^{~\nu}(x)=\delta^{\nu}_{~\mu}~,\nonumber \\&
    e^{a}_{~\mu}(x) e_{b}^{~\nu}(x)=\delta^{a}_{~b}~.
\end{align}
The tetrad basis in not arbitrary, but it is expected to produce the proper metric tensor according to eq.(\ref{tetradprop1}). The rising and lowering of the indices is carried out with the metric tensor for general coordinate and with the flat spacetime metric for the local inertial frames. The tetrad field is a set of four covariant vector fields (not a single rank two tensor), and reads 
\begin{align}
    e^{a}_{~\mu} (x)=\lbrace  e^{0}_{~\mu} (x), e^{1}_{~\mu} (x), e^{2}_{~\mu} (x), e^{3}_{~\mu} (x)  \rbrace ~.
\end{align}
The tetrad fields transform in general coordinate system as 
\begin{align}
    e^{a}_{~\mu} (x)\rightarrow \tilde{e}^{a}_{~\mu} (\tilde{x})=\frac{\partial x^\nu}{\partial\tilde{x}^\mu}e^{b}_{~\nu} (x)
\end{align}
and under a Lorentz transformation in local inertial frame as
\begin{align}
    e^{a}_{~\mu} (x)\rightarrow \tilde{e}^{a}_{~\mu} (\tilde{x})=\Lambda^a_{~b}(x)e^b_{~\mu}(x)~.
\end{align}
Since it is impossible to express spinors in the context of general relativity, tetrads play a crucial role to describe spin-1/2 particles in the presence of gravitational fields. It is also important to note that any tetrad representation of a specific metric tensor is not unique, and hence, different tetrad fields will lead to the same metric tensor if they are related by local Lorentz transformations \cite{lanzagorta,felice}. In our context, we shall consider the metric tensor for slowly rotating black hole spacetime. 
\noindent The slowly rotating black hole spacetime is defined by the line element 
\begin{align}
    ds^{2}=&-f(r) dt^{2}+\frac{1}{f(r)} dr^{2}+r^{2}(d\theta^{2}+\sin^{2}{\theta} ~d\phi^{2})\nonumber \\ &-\frac{2 a r_{s}}{r} \sin^{2}{\theta} ~dt ~d\phi~.
\end{align}
Here, $f(r)=1-\frac{r_s}{r}$ with $r_s = 2G_0 M$ being the Schwarzschild radius and $a$ being the rotation parameter, related to the spin angular momentum of the black hole. Here we have assumed `$a$' to be very small compared to the mass of the black hole. For the Schwarzschild spacetime where the line element is diagonal, the tetrads are calculated using the following diagonal ansatz \cite{lanzagorta}
\begin{align}
   & e_a^{~t}=(e_0^{~t},0,0,0)~,  e_a^{~r}=(0,e_1^{~r},0,0)~, \nonumber \\ & e_a^{~\theta}=(0,0,e_2^{~\theta},0)~, e_a^{~\phi}=(0,0,0,e_3^{~\phi})~.
\end{align}
In the rotating case the above ansatz will not work, and we need to modify the ansatz to have the tetrads for the non-diagonal line element as
\begin{align}
   & e_a^{~t}=(e_0^{~t},0,0,e_0^{~\phi})~,  e_a^{~r}=(0,e_1^{~r},0,0)~,\nonumber \\ & e_a^{~\theta}=(0,0,e_2^{~\theta},0)~, e_a^{~\phi}=(e_3^{~t},0,0,e_3^{~\phi})~.
\end{align}
Now using the properties of the tetrads in eq.(\ref{tetradprop1}) and eq.(\ref{tetradprop2}), we have obtained the non-zero component of the tetrads in the slowly rotating black hole spacetime as 
\begin{align}
       & e_{0}^{~t} =\frac{1}{\sqrt{f(r)}},~~~
        e_1 ^{~r}=\sqrt{f(r)},~~
        e_2 ^{~\theta}=\frac{1}{r}\nonumber \\ &
        e_3 ^{~t}=-\frac{a r_s}{r^2 f(r)},~~
        e_3 ^{~\phi}=\frac{1}{r}~.
\end{align}
To derive this, we have retained terms only up to linear order in the rotation parameter $a$.

\section{co-moving tetrads for slowly rotating spacetime}\label{cotetrad}
\noindent  In this section, we shall discuss the tetrads that move along the inertial observer. This set of tetrads is called the \textit{co-moving tetrads}. In this frame, the accelerated observers are instantaneously at rest \cite{plyatsko,bolos1,bolos2,schutz} and the timelike leg of the tetrads is identified with the four-velocity $\tilde{u}^{\mu}$ of the particle (co-moving observer), which means that the co-moving tetrads satisfy the following conditions :
\begin{align}\label{comovingprop}
    e_{0}^{~\mu}=\tilde{u}^{\mu},~
    e^{a}_{~\mu}\tilde{u}^{\mu}=0~,~~ a=1,2,3 
\end{align}
where the four-velocity $\tilde{u}^{\mu}$ follows the normalization condition
\begin{equation}\label{norm}
    \tilde{u}_{\mu}\tilde{u}^{\mu}=-1~.
\end{equation}
Now using these properties along with eq.(\ref{tetradprop1}) and eq.(\ref{tetradprop2}), we shall find the co-moving tetrads for the slowly rotating black hole spacetime. As we have mentioned in the earlier section, we have taken the ansatz to find the co-moving tetrads as follows
    \begin{align}
         &e_0^{~\mu}=(\tilde{u}^t,\tilde{u}^r,0,\tilde{u}^\phi)~,e_1^{~\mu}=(e_1^{~t},e_1^{~r},e_1^{~\theta},e_1^{~\phi})~,\nonumber \\&  e_2^{~\mu}=(e_2^{~t},e_2^{~r},e_2^{~\theta},e_2^{~\phi})~,e_3^{~\mu}=(e_3^{~t},e_3^{~r},e_3^{~\theta},e_3^{~\phi})~.
    \end{align}
Using the properties of co-moving tetrads and the normalization condition of four velocity discussed earlier, we formulate a system of algebraic equations whose solutions yield the co-moving tetrads in the spacetime of slowly rotating black hole. The equations we obtained up to $\mathcal{O}(a)$ are as follows
\begin{widetext}
\begin{align}
    &\frac{1}{f}(\tilde{u}^r)^2+r^2 (\tilde{u}^\phi)^2-\frac{2ar_s}{r}\tilde{u}^t\tilde{u}^\phi-f (\tilde{u}^t)^2=-1~,~~
    (e_1^{~t})^2+(e_2^{~t})^2+(e_3^{~t})^2-(\tilde{u}^{t})^2=-\frac{1}{f},\nonumber \\&
    (e_1^{~r})^2+(e_2^{~r})^2+(e_3^{~r})^2-(\tilde{u}^{r})^2=f~,~~
    (e_1^{~\theta})^2+(e_2^{~\theta})^2+(e_3^{~\theta})^2=\frac{1}{r^2},\nonumber\\&
    (e_1^{~\phi})^2+(e_2^{~\phi})^2+(e_3^{~\phi})^2-(\tilde{u}^{\phi})^2=\frac{1}{r^2}~,~~
    e_1^{~t}e_1^{~\phi}+e_2^{~t}e_2^{~\phi}+e_3^{~t}e_3^{~\phi}-\tilde{u}^t \tilde{u}^\phi=-\frac{ar_s}{r^3f},\nonumber \\&
    \frac{1}{f}\tilde{u}^re_1^{~r}+r^2 \tilde{u}^\phi e_1^{~\phi}-\frac{ar_s}{r}(\tilde{u}^t e_1^{~\phi}+\tilde{u}^\phi e_1^{~t})-f\tilde{u}^t e_1^{~t}=0~,~~
     \frac{1}{f}\tilde{u}^r e_2^{~r}+r^2 \tilde{u}^\phi e_2^{~\phi}-f\tilde{u}^t e_2^{~t}=0,\nonumber\\&
      \frac{1}{f}\tilde{u}^r e_3^{~r}+r^2 \tilde{u}^\phi e_3^{~\phi}-f\tilde{u}^t e_3^{~t}-\frac{ar_s}{r}(\tilde{u}^t e_3^{~\phi}+\tilde{u}^\phi e_3^{~t})=0~,~~
\frac{1}{f}e_1^{~r}e_2^{~r}+r^2 (e_1^{~\theta}e_2^{~\theta}+ e_1^{~\phi}e_2^{~\phi})=0,\nonumber\\&
\frac{1}{f}e_1^{~r}e_3^{~r}+r^2 e_1^{~\phi}e_3^{~\phi}-fe_1^{~t}e_1^{~t}-\frac{ar_s}{r}e_1^{~\phi}e_3^{~t}=0~,~~
\frac{1}{f}e_2^{~r}e_3^{~r}+r^2 (e_2^{~\theta}e_3^{~\theta}+e_2^{~\phi}e_3^{~\phi})-fe_2^{~t}e_3^{~t}=0,\nonumber\\&
\frac{1}{f}(e_1^{~r})^2+r^2 (e_1^{~\phi})^2=1~,~~
\frac{1}{f}(e_3^{~r})^2+r^2(e_3^{~\phi})^2-f(e_3^{~t})^2-\frac{2ar_s}{r}e_3^{~t}e_3^{~\phi}=0~.
\end{align}
\end{widetext}
\noindent Solving these equations, we have obtained the co-moving tetrads for slowly rotating spacetime and found the non-zero components up to $\mathcal{O}(a)$ to be
\begin{widetext}
\begin{align}
       & e_{0}^{~t} = u^t-\frac{a r_s u^\phi}{f r}~,\
        e_0 ^{~r}=u^r-\frac{a f r_s u^t u^\phi}{ r u^r}~,\
        e_0 ^{~\phi}= u^\phi+\frac{a r_s u^t}{r^3}~, 
        e_1 ^{~t}=-\frac{a p r_s \left(\frac{p^2 f}{r^2}+\frac{{(u^\phi)}^2}{p^2 f}\right)}{u^r}~,\nonumber \\&
        e_3 ^{~t}=\sqrt{\frac{f {(u^t)}^2-1}{f}}-\frac{a r_s u^t u^\phi }{b f r}~,
        e_1 ^{~r}=r u^\phi  \sqrt{\frac{f}{f {(u^t)}^2-1}}-\frac{a p f r_s u^t}{r^2}~,
        e_3 ^{~r}=u^r u^t \sqrt{\frac{f}{f {(u^t)}^2-1}}~,
        e_2 ^{~\theta}=\frac{1}{r}~,\nonumber \\&
        e_1 ^{~\phi}=-\frac{u^r}{r \sqrt{f \left(f {(u^t)}^2-1\right)}}-\frac{a p f r_s u^t u^\phi}{r^2 u^r}~,
        e_3 ^{~\phi}=u^t u^\phi \sqrt{\frac{f}{f {u^t}^2-1}}~.
\end{align}
\end{widetext}
Here, $p$ is defined as $p=\sqrt{\frac{f(r)(u^t)^2 -1}{f(r)}}$ and $u^{\mu}$ represents the four-velocity of the particle in Schwarzschild spacetime, which also follows the normalization condition $u_{\mu}u^{\mu}=-1$. In the $a\rightarrow 0$ limit, these tetrads reduce to the co-moving tetrads in Schwarzschild spacetime \cite{lanzagorta}. We shall use these tetrads to compute the force due to the curvature-spin coupling in a slowly rotating spacetime.

\section{spin-curvature forces in general relativity}\label{forcegeneral}
\noindent In this section, we shall briefly review the Dirac equation in curved spacetime and the force arising due to spin-curvature coupling \cite{audretsch,cianfrani1,cianfrani2,hafner} up to the linear order of $\hbar$. To describe the Dirac spinors in curved spacetime, we shall use the WKB approximation method, assuming that a four-spinor $\Psi (x)$ can be expressed by a local phase factor $\mathcal{S}(x)$ and a power series of $\hbar$ as \cite{audretsch}
\begin{equation}\label{diracansatz}
\Psi(x)=e^{i \mathcal{S}(x)/\hbar}\sum_{n=0}^{\infty}(-i\hbar)^{n}\psi_{n}(x)
\end{equation}

\noindent with $\psi_{n}$ being the four-spinors. The Dirac equation in curved spacetime reads
\begin{equation}\label{diraccurved}
    (i\gamma^{\mu}(x)\mathcal{D}_{\mu}-m)\Psi(x)=0~.
\end{equation}
Here, $\gamma^{\mu}(x)$ are the gamma matrices in the general coordinate system and $\mathcal{D}_{\mu}$ denotes the covariant derivative with 
\begin{equation}
    \mathcal{D}_{\mu}\Psi\equiv (\partial_{\mu}-\Gamma_{\mu})\Psi
\end{equation}
where the spinorial affine connection $\Gamma_{\mu}$ has the form as follows
\begin{equation}
    \Gamma_{\mu}(x)=\frac{1}{8}\omega_{\mu}^{~ab}(x)[\gamma_{a},\gamma_{b}]~.
\end{equation}
Here $\omega_{\mu}^{~~ab}(x)$ are the connection one-forms defined as
\begin{equation}
    \omega_{\mu ab}(x)\equiv \eta_{ac}e^{c}_{~\nu}(\partial_{\mu}e_{b}^{~\nu}+\Gamma^{\nu}_{~\mu \lambda}e_{b}^{~\lambda})~.
\end{equation}
Using the expansion in eq.(\ref{diracansatz}) and substituting in eq.(\ref{diraccurved}), we obtain the Dirac equations for the zeroth and the first order in $\hbar$ as \footnote{Here $\mathcal{S}(x)$ is the slowest varying component and the spinors $\psi_n (x)$ are the fastest varying components of the field. Hence, the expansion of the phase is not needed.}
\begin{align}
    &(\gamma^{\beta}\partial_{\beta}\mathcal{S}(x)+m)\psi_{0}(x)=0 \label{h0}\\&
    (\gamma^{\beta}\partial_{\beta}\mathcal{S}(x)+m)\psi_{1}(x)=-\gamma^{\beta}\mathcal{D}_{\beta}\psi_{0}(x) \label{h1}~.
\end{align}
The gamma matrices follow the usual anti-commutation relation
\begin{equation}
    \lbrace \gamma^\alpha (x), \gamma^\mu (x)\rbrace = 2 g^{\alpha \mu} (x)~.
\end{equation}
The Dirac equation at the order $\hbar^0$ is a homogeneous system of four algebraic equations, which can have a non-trivial solution only if the condition
\begin{equation}\label{solcond}
    det\left(\gamma^{\beta}\partial_{\beta}\mathcal{S}(x)+m\right)=0
\end{equation}
is met.
This leads to the Hamilton-Jacobi equation for a relativistic spinless particle of mass $m$
\begin{equation}
    \partial_{\beta}\mathcal{S}(x) \partial^{\beta}\mathcal{S}(x)=-m^2 ~.
\end{equation}
At $\hbar^0$ order, it is convenient to introduce $\mathcal{S}(x)$ as the classical action for a spinless particle in a gravitational field
\begin{equation}
    \mathcal{S}(x)=\int p_\beta dx^\beta
\end{equation}
where, $p_\beta$ is the four-momentum of the particle, defined as $p_\beta \equiv \partial_\beta \mathcal{S}(x)$ and consequently the four-velocity of the particle can be written as $\tilde{u}_\beta \equiv \frac{1}{m}\partial_\beta \mathcal{S}(x)$. Therefore, in this order, the quantum mechanical phase $\mathcal{S}(x)$ can be identified by considering the Dirac particle as a spinless classical particle moving with momentum $p^\beta$ in curved spacetime \cite{alsing, stodolsky}. The two positive energy solutions\footnote{For this solution, the Dirac-Pauli representation of gamma matrices are used.} for eq.(\ref{h0}) can be written as 
\begin{align}
    &\psi_{01}(x)=\sqrt{\frac{e^0_{~\nu}(x)p^\nu +m}{2m}}\begin{pmatrix}
        1 \\ \\ 0 \\ \\
        \frac{e^3_{~\nu}(x)p^\nu}{e^0_{~\nu}(x)p^\nu +m}\\ \\
        \frac{e^1_{~\nu}(x)p^\nu+ie^2_{~\nu}(x)p^\nu}{e^0_{~\nu}(x)p^\nu +m}
    \end{pmatrix} \nonumber \\\nonumber \\&
    \psi_{02}(x)=\sqrt{\frac{e^0_{~\nu}(x)p^\nu +m}{2m}}\begin{pmatrix}
        0 \\ \\ 1 \\ \\
        \frac{e^1_{~\nu}(x)p^\nu-ie^2_{~\nu}(x)p^\nu}{e^0_{~\nu}(x)p^\nu +m}\\ \\
        \frac{e^3_{~\nu}(x)p^\nu}{e^0_{~\nu}(x)p^\nu +m}
    \end{pmatrix}~,
\end{align}
see \cite{audretsch,lanzagorta}.
Therefore, the general solution for eq.(\ref{h0}) is given by
\begin{equation}
    \psi_0 = c_1 (x) \psi_{01}(x)+c_2 (x) \psi_{02}(x)
\end{equation}
where the coefficients $c_1 (x)$ and $c_2 (x)$ are complex scalars. The condition eq.(\ref{solcond}) for this non-trivial solution of $\psi_0(x)$ restricts the solution of eq.(\ref{h1}) which is a linear inhomogeneous algebraic equation for $\psi_1(x)$. Hence, the condition for having non-trivial solution for $\psi_1$ is that all the solutions of the homogeneous equation (\ref{h0}) have to be orthogonal to the inhomogeneity 
\begin{align}
    &\bar{\psi}_{01}(x)\gamma^\beta \mathcal{D}_{\beta}\psi_0 (x) =0 \nonumber \\&
    \bar{\psi}_{02}(x)\gamma^\beta \mathcal{D}_{\beta}\psi_0 (x) =0~.
\end{align} 

\noindent As in this framework the curvature and spin are coupled in a non-trivial way, one needs to study the influence of the spin on the trajectories of the particle using the Gordon decomposition \cite{gordon,itzykson,bjorken} of the Dirac probability current which is given by 
\begin{equation}
    j^\mu = \bar{\Psi}\gamma^\mu \Psi 
\end{equation}
and which satisfies the continuity equation
\begin{equation}
    \mathcal{D}_\mu j^\mu =0~.
\end{equation}
Applying Gordon decomposition, this current can be expressed as the sum of convection current and magnetization current 
\begin{equation}
    j^\mu = j^\mu_C +j^\mu_M
\end{equation}
with
\begin{align}
    & j^\mu_C =\frac{\hbar}{2mi}\biggr(\mathcal{D}^\mu \bar{\Psi} \Psi - \bar{\Psi}\mathcal{D}^\mu \Psi \biggr) \label{convection}\\&
    j^\mu_M = \frac{\hbar}{2m}\mathcal{D}_\nu \left(\bar{\Psi} \sigma^{\mu \nu}\Psi \right)
\end{align}
where $\sigma^{\alpha \beta}\equiv \frac{i}{2}[\gamma^\alpha,\gamma^\beta]$. After inserting the WKB expansion in eq.(\ref{diracansatz}) up to the first order, eq.(\ref{convection}) takes the following form
\begin{align}
    j^\mu_C =& \tilde{u}^\mu \biggr[ \wp^2 -i\hbar \left(\bar{\psi}_0 \psi_1 - \bar{\psi}_1 \psi_0 \right) \biggr]\nonumber\\&-\frac{i\hbar}{2m}\biggr(\psi_0 \mathcal{D}^\mu \bar{\psi}_0 - \bar{\psi}_0 \mathcal{D}^\mu \psi_0 \biggr)+\mathcal{O}(\hbar^2)~,
\end{align}
where $\wp^2 = \bar{\psi}_0 \psi_0 $. As discussed in \cite{bjorken}, the convection current defines the probability flow of a particle moving with four-velocity $\tilde{v}^\alpha$ 
\begin{equation}
    j^\mu_C \sim \tilde{v}^\mu 
\end{equation}
where $\tilde{v}^\alpha$ satisfies the normalization $\tilde{v}_\alpha \tilde{v}^\alpha =-1$.
After normalizing the convection current, one can reach the expression 
\begin{align}
    \tilde{v}^\mu =& \tilde{u}^\mu -\frac{i\hbar \tilde{u}^\mu}{\wp^2} \left(\bar{\psi}_0 \psi_1 - \bar{\psi}_1 \psi_0 \right)\nonumber\\& -\frac{i\hbar}{2m}\biggr(\psi_0 \mathcal{D}^\mu \bar{\psi}_0 - \bar{\psi}_0 \mathcal{D}^\mu \psi_0 \biggr)+\mathcal{O}(\hbar^2)~,
\end{align}
which in turn gives the correction to the geodesic four-velocity $\tilde{u}^\alpha$ due to the curvature-spin interaction, and up to $\mathcal{O}(\hbar)$ the correction evaluates to
\begin{equation}
    \delta \tilde{u}^\mu = \frac{\hbar \tilde{u}^\mu}{i\wp^2} \left(\bar{\psi}_0 \psi_1 - \bar{\psi}_1 \psi_0 \right) -\frac{i\hbar}{2m}\biggr(\psi_0 \mathcal{D}^\mu \bar{\psi}_0 - \bar{\psi}_0 \mathcal{D}^\mu \psi_0 \biggr)~.
\end{equation}
This deviation from geodesics occurs due to a force arising from the spin-curvature coupling \cite{audretsch}. The formal expression for this spin-curvature coupling force is therefore given as 
\begin{align}\label{spincurvatureforce}
    f^\alpha &=\frac{\hbar}{4} g^{\alpha \mu} \tilde{u}^\beta R_{\mu \beta \gamma \delta} e^{~~ \gamma}_{a} e^{~~\delta}_{b} \bar{\psi}_0 \sigma ^{ab} \psi_0 \nonumber \\
    &+\frac{m \hbar}{\wp^2 i} g^{\alpha \mu} \tilde{u}^\beta \bigg(\mathcal{D}_\beta \left(\left(\bar{\psi}_0 \psi_1-\bar{\psi}_1 \psi_0\right)\tilde{u}_\mu \right)\nonumber \\
    &-\mathcal{D}_\mu \left(\left(\bar{\psi}_0 \psi_1-\bar{\psi}_1 \psi_0\right)\tilde{u}_\beta \right )\bigg)
\end{align}
up to linear order of $\hbar$. This constitutes an important result of the present analysis. In \cite{lanzagorta,audretsch}, the second term which is of $\mathcal{O}(\hbar)$ was not taken into account. In our analysis, however, we retain both terms in the force expression to capture the complete $\mathcal{O}(\hbar)$ correction to the geodesic deviation arising from the spin-curvature coupling force.  In this article, we shall focus on evaluating and analyzing the first term of the above expression of the spin-curvature coupling force in slowly rotating black hole spacetime. In this expression eq.(\ref{spincurvatureforce}), the first term manifestly shows that the force arises due to the interaction between the curvature of spacetime, defined by $R_{\mu \beta \gamma \delta}$, and the quantum spin of the particle.

\section{spin-curvature coupling force in slowly rotating black hole spacetime}\label{forcerotating}
\noindent In this section, we have presented the force components for the slowly rotating black hole spacetime, where we have considered a test particle following the geodesic motion in the equatorial plane ($\theta = \frac{\pi}{2}$) with four-velocity $\tilde{u}^\mu$ \footnote{In equatorial plane, the $\theta$ component of the four-velocity $\tilde{u}^\theta =0$.}. Using eq.(\ref{spincurvatureforce}), we have computed the force components up to the linear order of the rotation parameter ($a$) as follows:
\begin{widetext}
\begin{align}\label{forcecomponents}
    &f^t=\left(\frac{3 \hbar r_s  u^r u^\phi }{4 f r^2}+\frac{3 a \hbar r_s u^t \left(r r_s  (u^\phi)^2+(u^r)^2\right)}{4 r^4 u^r}\right) \bar{\psi}_{0} \sigma^{13}\psi_{0}~,\\&
f^r=\left(\frac{3 f \hbar r_s  u^t u^\phi}{4 r^2}+\frac{3 a \hbar r_s   \left(f^2 (u^t)^2+r^2 (u^\phi)^2\right)}{4 r^4} \right)\bar{\psi}_{0} \sigma^{13}\psi_{0}~,\\&
 f^\theta = -\frac{3 \hbar r_s  u^r u^\phi}{4 r^3 \sqrt{f \left(f {(u^t)}^2-1\right)}}\bar{\psi}_{0} \sigma^{12}\psi_{0}-\frac{3 f \hbar r_s  u^t {(u^\phi)}^2}{4 r^2 \sqrt{f \left(f {(u^t)}^2-1\right)}}\bar{\psi}_{0} \sigma^{23}\psi_{0}+\frac{ a \hbar r_s u^\phi \left(-3 f^2 {(u^t)}^2+f (u^t)^2+1\right)}{4 r^4 \sqrt{f \left(f {(u^t)}^2-1\right)}}\bar{\psi}_{0} \sigma^{23}\psi_{0} \nonumber \\&~~~~~~~+ \Bigg( \frac{a \hbar r_s u^t \left(2 r {(u^r)}^2\right)}{4 f p r^6 u^r}+\frac{3 a \hbar r_s u^t \left(f r^2 r_s {(u^t)}^2 {(u^\phi)}^2-r^2 r_s {(u^\phi)}^2-r {(u^r)}^2\right)}{4 p r^6 u^r}\Bigg)\bar{\psi}_{0} \sigma^{12}\psi_{0}~,\\ \nonumber \\&
   f^\phi=\frac{3 a \hbar r_s  u^r u^\phi}{4 f r^4}\bar{\psi}_{0} \sigma^{13}\psi_{0}~.
\end{align}
\end{widetext}
Using the spinor $\psi_0$ in co-moving frame, the force components can be expressed in terms of spin components $s_1$, $s_2$ and $s_3$ as
\begin{widetext}
\begin{align}\label{forcecomponentspin}
    &f^t=\left(\frac{3 \hbar r_s  u^r u^\phi }{4 f r^2}+\frac{3 a \hbar r_s u^t \left(r r_s  (u^\phi)^2+(u^r)^2\right)}{4 r^4 u^r}\right) s_2 ~,\\&
f^r=\left(\frac{3 f \hbar r_s  u^t u^\phi}{4 r^2}+\frac{3 a \hbar r_s   \left(f^2 (u^t)^2+r^2 (u^\phi)^2\right)}{4 r^4} \right)s_2 ~,\\&
 f^\theta = -\frac{3 \hbar r_s  u^r u^\phi}{4 r^3 \sqrt{f \left(f {(u^t)}^2-1\right)}}s_3-\frac{3 f \hbar r_s  u^t {(u^\phi)}^2}{4 r^2 \sqrt{f \left(f {(u^t)}^2-1\right)}}s_1+\frac{ a \hbar r_s u^\phi \left(-3 f^2 {(u^t)}^2+f (u^t)^2+1\right)}{4 r^4 \sqrt{f \left(f {(u^t)}^2-1\right)}}s_1 \nonumber \\&~~~~~~~+ \Bigg( \frac{a \hbar r_s u^t \left(2 r {(u^r)}^2\right)}{4 f p r^6 u^r}+\frac{3 a \hbar r_s u^t \left(f r^2 r_s {(u^t)}^2 {(u^\phi)}^2-r^2 r_s {(u^\phi)}^2-r {(u^r)}^2\right)}{4 p r^6 u^r}\Bigg)s_3 ~,\\ \nonumber \\&
   f^\phi=\frac{3 a \hbar r_s  u^r u^\phi}{4 f r^4}s_2~.
\end{align}
\end{widetext}
In $a\rightarrow 0$ limit, these results agree with the force components calculated in Schwarzschild spacetime \cite{lanzagorta} and the spin components for three qubit states in the equatorial plane are given in \cite{lanzagorta}. \\
Now, the magnitude of the spin-curvature force up to $\mathcal{O}(a)$ can be obtained as
\begin{widetext}
\begin{align}\label{fscsq}
    |f_{sc}|^2 &=g_{\mu\nu} f^\mu f^\nu \nonumber \\ 
    &=\frac{3\hbar^2 r_s^2 u^t (u^\phi)^3 \biggr\lbrace 3f r^2 u^t u^\phi +2 a \biggr(3f^2 (u^t)^2 - 2 f (u^t)^2-1 \biggr) \biggr\rbrace}{16r^4\biggr( f (u^t)^2-1\biggr)} s_1^2 \nonumber \\ & + \frac{9\hbar^2 r_s^2 u^\phi \biggr\lbrace r^2 u^\phi \left( f^2 (u^t)^2-(u^r)^2 \right)+2af u^t \left( f^2 (u^t)^2-(u^r)^2 +r^2 f (u^\phi)^2\right) \biggr\rbrace}{16f r^6} s_2^2 \nonumber \\& + \frac{3\hbar^2 r_s^2 u^\phi \biggr\lbrace 3fr^2 p(u^r)^2 u^\phi-2afpu^t \biggr((2-3f)(u^r)^2+3f^2p^2r r_s (u^\phi)^2 \biggr) \biggr\rbrace }{16f^3 r^6 p^3} s_3^2 \nonumber \\& - \frac{3\hbar^2 r_s^2 (u^\phi)^2 \biggr\lbrace -3fpr^2u^t u^\phi (u^r)^2 +3arr_s f^3 p^3 (u^t)^2 (u^\phi)^2 +ap(u^r)^2(1-3f^2 (u^t)^2) \biggr\rbrace}{8 f^2 p^3 r^5 u^r}s_1 s_3 ~. 
\end{align}
\end{widetext}
Now, we consider the circular orbit condition and compute the magnitude of the force. For the circular orbit in slowly rotating spacetime $\tilde{u}^r|_{r=R} =0$ with $R$ being the radius of the circular orbit. We shall use the expressions for the four-velocity as $u^t=\frac{K}{f(R)}$, $(u^r)^2=\frac{a f r_s u^t u^\phi}{R}$ and $u^\phi=\frac{J}{R^2}$, with \cite{lanzagorta}
\begin{align}
    J^2=\frac{1}{2}\frac{R~ r_s}{\left(1-\frac{3r_s}{2R}\right)} ~~, ~~
K=\frac{1- \frac{r_s}{R}}{\sqrt{\frac{R r_s}{1-\frac{3r_s}{2R}}}}.
\end{align}
From the expression in eq.(\ref{fscsq}), in the large $r$ limit, imposing the circular orbit conditions, we find the following behaviour of the force,
\begin{equation}
    |f_{sc}| \sim \mathcal{O} (R^{-\frac{7}{2}})+\mathcal{O} (R^{-\frac{17}{4}})\sqrt{a}+\mathcal{O} (R^{-4}) a.
\end{equation}
The fact we find here is the radial fall of the force magnitude is more rapid than Newtonian Gravitational force and rotating corrections are even more rapidly falling in comparison with the Schwarzchild black hole spin-curvature force magnitude.

\section{Discussion and future directions}\label{conclusion}
\noindent In this article, we have investigated the relativistic quantum force arising from the interaction between the quantum particle's intrinsic spin and the curvature of the spacetime geometry. This force causes the massive spin-1/2 quantum particles to deviate from the geodesic motion. We have considered massive spin-1/2 particles propagating in the background of a slowly rotating black hole spacetime. Unlike in flat spacetime, where the Dirac equation can be expressed using the constant gamma matrices; in curved spacetime, it requires more general structure. In particular, here the spin connection terms arise and consequently, the ordinary partial derivative in the flat space Dirac equation is replaced by more general covariant derivative.  
Further, since the $\psi (x)$ is a spinor field, we need to invoke the concept of tetrads in this context, which provide the spacetime dependent gamma matrices in the Dirac equation in the given black hole background geometry. \\
\noindent In the present analysis, we have adopted the concept of co-moving tetrads, in which the timelike tetrad leg is identified with the four velocity of the massive particle. This adoption is convenient because we are seeking the spin dynamics from the point of view of an observer locally moving with the particle.
We have constructed the co-moving tetrads in a slowly rotating black hole spacetime, extending the formalism used to compute the same for Schwarzschild geometry \cite{lanzagorta}. The importance of considering this geometry as the background is that rotation brings in additional effects such as frame dragging since the metric has a cross-term between the coordinate time ($t$) and the azimuthal angular coordinate $\phi$. The calculations have been performed up to linear order in the rotation parameter ($a$).
\noindent We have reestablished the expression for the spin-curvature coupling force in section \ref{forcegeneral}. Unlike the treatment in \cite{lanzagorta,audretsch}, where the second term of the force was not taken into account, we have included it to capture the complete $\mathcal{O}(\hbar)$ correction to the spin-curvature coupling force.
Employing the co-moving tetrads mentioned in section \ref{cotetrad}, we have obtained the components of the spin-curvature coupling force again up to  first order in $a$. This force creates a correction to the geodesic four velocity of the quantum spin-$1/2$ particle, which means that the particle deviates from its geodesics in the slowly rotating black hole spacetime geometry. We have used the Gordon decomposition for the Dirac probability four current to study the effect of the intrinsic spin on the trajectories of the particles.
Then we have computed the magnitude of this relativistic quantum force in the given spacetime upto $\mathcal{O}(a)$. We have extended our analysis for a special case, which is the circular orbit of the particles, where the radial component of the four velocity vanishes. In the circular orbit condition, we have studied the behaviour of the force magnitude for a particular orbit radius with $R$. Comparing our results with the same for Schwarzschild geometry mentioned in \cite{lanzagorta}, we found that the rotation parameter modifies the force magnitude and introduces an additional rotation dependent term that falls with radial distance much rapidly than the rotation independent term in the Schwarzschild case. Since the contribution to the relativistic quantum force coming from the rotation parameter of the spacetime geometry decays more rapidly even than the non-rotating counter part, this result gives firm footing to the remark that the spin-curvature coupling force is much weaker than the Newtonian gravitational force for the spinless particles \cite{lanzagorta}.\\
\noindent The present analysis in our article also opens up several directions for future investigations. The natural first extension is to go beyond the slowly rotating black hole spacetime and determine the spin-curvature coupling force in the full Kerr black hole geometry, and the radial power law behaviour of the magnitude of the force and the dependence on the black hole rotation parameter. Next, we can go to the near extremal limits of the Kerr black hole, where this force might get significantly modified due to strong curvature and frame dragging effects. As the spin-curvature coupling force modifies the trajectory of the Dirac particle in Kerr black hole spacetime, one can analyze the corrections to Wigner rotations calculated earlier in the classical geodesic approximation. This extension will give us a connection between the spin curvature coupling effect and the propagation of quantum information in strongly rotating gravitational background geometry. It would also be interesting to generalize our current analysis to rotating-charged black hole (Kerr-Neumann) case and other spacetimes with cosmological constant, and see how the spin-curvature coupling force behaviour differs in comparison to our current results. We hope to return to these questions in future.

\section{Acknowledgement} 
\noindent The work of S.M. was supported by the Advanced Postdoctoral Research Programme (APRP) of S. N. Bose National Centre for Basic Sciences.
\section{Appendix : Connection one-forms for Dirac field in co-moving tetrad basis}
\noindent In this Appendix, we report the expressions for connection one forms for Dirac fields in the slowly rotating black hole background in co-moving tetrad basis, which in turn generate the spin-connections in Dirac equation in curved spacetime. The components are computed using the formal expression for connection one-forms \cite{lanzagorta}
\begin{align}
    \omega_{\mu a b}=\eta_{ac}~e_{\nu}^{~c}\biggr(\partial_{\mu}e_{b}^{~\nu}+\Gamma^{\nu}_{~\mu\lambda}e_{b}^{~\lambda} \biggr)~.
\end{align}
Here, the affine connections are calculated in the slowly rotating black hole spacetime geometry upto $\mathcal{O}(a)$ in an equatorial plane $\theta=\pi/2$, and the tetrads are used in co-moving frame, which are mentioned in the earlier section. The components of $\omega_{\mu a b}$ read
\begin{widetext}
\begin{align*}
    &\omega_{t02}=\omega_{r02}=\omega_{\theta 01}=\omega_{\theta 03}=\omega_{\phi 02}=\omega_{t12}=\omega_{t23}=\omega_{r12}=\omega_{r23}=\omega_{\theta 12}=\omega_{\phi 12}=\omega_{\theta 13}=\omega_{\phi 23}=\omega_{\theta 11}=\omega_{\phi 11}=0,\\&
    \omega_{t11}=\omega_{t22}=\omega_{r22}=\omega_{\theta 22}=\omega_{\phi 22}=\omega_{t33}=\omega_{\theta33}=\omega_{\phi33}=0,\\&
\omega_{t01}=-\frac{r_s u^t u^\phi}{2 r p}+\frac{a p r_s}{2r^3},~ \omega_{t03}=-\frac{r_s u^r}{2 r^2 f p}-\frac{a r_s^2 f p u^t u^\phi}{2 r^3 u^r},~   \omega_{\phi 13}=-u^t f-\frac{a r_s u^\phi}{2r},~\omega_{\theta 23}=\frac{u^r u^t}{p},\\&
 \omega_{r01}=\bigg(-f u^t \partial_r e^{~t}_1+\frac{\tilde{u^r}}{f}\partial_r e^{~r}_1+r^2 u^\phi \partial_r e^{~\phi}_1\bigg)
    +\frac{(2r-r_s)u^r u^\phi}{2 f r^2 p}- \frac{a r_s u^t}{2 f p r^4 u^r}\bigg(f^2 p^4 r_s-3r (u^r)^2 +p^2 r_s (u^r)^2 +2 r^2 (r_s-f^2 p^2 r){u^\phi}^2\bigg), \\&
 \omega_{r03}=\bigg(-f u^t \partial_r e^{~t}_3+\frac{\tilde{u^r}}{f}\partial_r e^{~r}_3+r^2 u^\phi \partial_r e^{~~\phi}_3\bigg)
     + \frac{u^t}{2 p r^2}\left(p^2 r_s+\frac{rs {(u^r)}^2}{p^2}-2r^3 {(u^\phi)}^2 \right)
    -\frac{a r_s }{2 f p r^3}\left(p^2 r+(3 f r + 2 r_s) {(u^t)}^2\right)u^\phi,\\&
\omega_{\theta 02}=-u^r+\frac{a f r_s u^t u^\phi}{r u^r},~ \omega_{\phi 01}=fp+\frac{a r_s u^t u^\phi}{2rp},~ \omega_{\phi 03}=a\left(\frac{rs u^r(1+2 f {(u^t)}^2)}{2 f p r^2}+\frac{f r_s {(u^t)}^2 {(u^\phi)}^2}{p u^r}\right),\\&
 \omega_{t13}=\frac{r_s u^\phi}{2 r}-\frac{a r_s u^t f}{2 r^3},~\omega_{r11}=(e_t^{~1}\partial_r e_1^{~t}+e_r^{~1}\partial_r e_1^{~r}+e_\phi^{~1}\partial_r e_1^{~\phi})+\frac{2(u^r)^2-rr_s (u^\phi)^2}{2rf^2p^2},\\& \omega_{r13}=(e_t^{~1}\partial_r e_3^{~t}+e_r^{~1}\partial_r e_3^{~r}+e_\phi^{~1}\partial_r e_3^{~\phi})+\frac{(r_s-2r)u^r u^t u^\phi}{2rp^2 f^2}+\frac{a r_s}{2fr^4u^r}\bigg(-(p^2 f)^2 r_s
    -(r-r_s {u^t}^2) {u^r}^2+r^2 \left(2f^2 r {u^t}^2-rs)\right){u^\phi}^2\bigg),\\&
     \omega_{r33}=(e_t^{~3}\partial_r e_3^{~t}+e_r^{~3}\partial_r e_3^{~r}+e_\phi^{~3}\partial_r e_3^{~\phi})+\frac{2f^2r^4(u^t)^2 (u^\phi)^2-rr_s(1+(u^t)^2((u^r)^2+f(f(u^t)^2 -2)))}{2f^3p^2r^3}+\frac{3ar_su^tu^\phi}{r^2}~.   
\end{align*}

\end{widetext}

\end{document}